# A Survey of Real-Time Social-Based Traffic Detection

Hashim Abu-gellban
Department of Computer Science
Texas Tech University
Hashim.gellban@ttu.edu

*Abstract*— Online traffic news web sites do not always announce traffic events in areas in real-time. There is a capability to employ text mining and machine learning techniques on the twitter stream to perform event detection, in order to develop a real-time traffic detection system. In this present survey paper, we will deliberate the current state-of-art techniques in detecting traffic events in real-time focusing on five papers [1, 2, 3, 4, 5]. Lastly, applying text mining techniques and SVM classifiers in paper [2] gave the best results (i.e. 95.75% accuracy and 95.8% $F_1$-score).

## I. INTRODUCTION

The purpose of this survey paper is to detect and analyze traffic events by processing users' tweets in real-time. These tweets belong to a certain area and are written in different languages like English, Italian, and Thai languages. Also, the purpose is to support traffic and city administrations for managing scheduled or unexpected events in the city, in order to provide the ability to integrate this system to other traffic sensors (e.g., loop detectors, cameras, and infrared cameras) and develop monitoring system (e.g. detecting traffic difficulties). Furthermore, it is to provide a low-cost wide coverage of the road network especially in those areas where traditional traffic sensors are missing, and develop a new Intelligent Transportation System (ITS).

Event detection from social networks analysis is a more challenging problem than event detection from traditional media (like blogs and emails). This problem is because the texts are well-formatted in traditional media while Status Update Messages "SUMs" (e.g. tweets information) are unstructured and irregular texts (i.e. it contains informal or abbreviated words, misspellings, or grammatical errors), incomplete source of information (i.e. it is very brief), and a huge amount of useless information required to be filtered (e.g. over 40% of all tweets is pointless [13]).

This survey covers pioneer systems called ITSs which detect traffic events from social networks. These systems have some of the following features: 2-class classification (i.e. non-traffic, and traffic), multi-class classification (e.g. non-traffic, congestion/crash traffic, and external events traffic), built based on SOA architecture, and traffic detection in tweets regarding a specific language. As tweets are unstructured and irregular with useless information, the dataset preprocessing using text mining techniques are important to apply. We will discuss the impacts of these techniques. Many classifications techniques were employed; whereas, the SVM classifier gave the best results. Cross-Validation method had been applied to evaluate the predictive models which were built by the classifiers. Generally, we will discuss and analyze the results to critique the techniques that had been used. These techniques affected the results positively or negatively.

The importance of this research is coming from the need of the audience to get traffic information in real-time, while there is still a delay in traffic detection using the current monitoring systems and traffic sensors. It is important to note here that the target audience of this research field is the online traffic news web sites, radio traffic news channels, police stations, and drivers.

Much research had been accomplished to extract useful information for event detection systems. These detection systems have two types: small-scale events and large-scale events. For small-scale event, it usually has a small number of Status Update Messages "SUMs" (i.e. shared user messages in social networks which have text and meta-information like timestamp and geographic coordinates), and belong to a precise geographic location in a short period, such as: traffic, car crashes, fires, or local manifestations. Literature showed different ways of detection and classification of tweets for the small-scale system. For instance, Abel et al. [9] analyzed small-scale events like fires by extracting features using NER from Twitter streams and emergency network information. While Agarwal et al. [10] detected fires in a factory from Twitter stream analysis using standard NLP techniques and a Naive Bayes classifier.

On the other hand, the large-scale event is a huge number of SUMs and it has a wider temporal and geographic coverage like earthquakes and storms. Bansal et al. [14] presented a text mining analysis system to show the hottest topics in tweets on a specific time and in their places with an explanation regarding why these topics were interesting. Furthermore, Chew et al. [11] classified tweets of H1N1 containing related keywords and hashtags. While Sakaki et al. [12] detected earthquakes and typhoons in tweets by filtering specific keywords and employing SVM. Whereas, Li et al. [6] developed a system called TEDAS to retrieve incident-related tweets (e.g. thunderstorms). This was accomplished by filtering keywords and classifying tweets based on their features, such as: number of followers of the user, and the number of retweets.

In section II, the methodologies and technologies for detecting real-time traffic events from Twitter data streams are described and categorized according to the tools, preprocessing of the datasets, classifiers (e.g. SVM), evaluation (e.g. k-fold Cross-Validation), and general discussion to analyze the results. Section III provides a conclusion. Finally, future work is discussed in section IV.

## II. REAL-TIME TRAFFIC DETECTION IN TWITTER

We studied five research papers related to traffic detection using Twitter [1, 2, 3, 4, 5]. TABLE I and TABLE II shows the tools and methodologies used in the five recent research papers and their results. Some of the information does not mention in some papers; therefore, we use (N/A) to any information that is not mentioned in the related paper. In this section, we will discuss the tools which have been used in these papers and the methodologies of preparing the training datasets. Also, we will deliberate classifications methods, the evaluations of the predictive classifier models, and the analysis of results.

### A. Tools

Twitter's API was used to crawl relevant traffic tweets [1, 2, 3, 4, 5]. This API was used to directly access and fetch the public stream of tweets with specific filtering according to date, time, geolocation, language, and keywords as shown in TABLE I. Kokkinogenis et al. [1], Schulz et al. [3], and Li et al. [4] filtered tweets according to English language; whereas, D'Andrea et al. [2] and Wanichayapong et al. [5] focused on Italian, and Thai languages, respectively. The authors of the above mentioned papers filtered tweets according to several different keywords to crawl the tweets in a way that would allow an easy classification of them. For instance, Kokkinogenis et al. [1] employed the keywords (i.e. holes, flood, and pavement) to get "Road Conditions" tweets. Whereas, D'Andrea et al. [2] used "traffico" (traffic), "coda" (queue), or "incidente" (crash) to filter traffic class label in Italian tweets. Moreover, D'Andrea et al. [2] used Twitter4J library. This library is to wrap the Twitter API, in order to be called through Java programs in ITS.

Kokkinogenis et al. [1], D'Andrea et al. [2], and Schulz et al. [3] used WEKA (Waikato Environment for Knowledge Analysis) [9] to classify the extracted features from tweets. They employed the LIBSVM (i.e. integrated software for Support Vector Machine "SVM" classification) which is available through WEKA. LIBSVM is required to be downloaded first; then, the classpath must be modified to access "libsvm.jar" which is included in LIBSVM distribution. Whereas, Li et al. [4] and Wanichayapong et al. [5] used their own programs to classify tweets using a linear regression model and syntactic analysis techniques, respectively.

### B. Preprocessing

Event detection from social networks analysis requires preparing the data before classifying. This is because SUMs are unstructured and very brief. Also, they contain irregular texts and some useless in-formation which is required to be filtered. Therefore, the authors [1, 2, 3, 4, 5] employed several text mining techniques to prepare the data in a more formal way in order to be efficiently classified to give good results. Some of these techniques were used in many of these papers while others were just employed in one paper. This survey provides some pros and cons of these techniques to explain the key reasons for good and unpleasant results.

TABLE I.
LIST OF LANGUAGES, TOOLS, CLASS LABELS, AND KEYWORDS.

| References | Language | Tools | | Class Label | Keywords |
|---|---|---|---|---|---|
| | | Crawling | WEKA | | |
| 1 | English | Twitter's API | Yes | 1. traffic and posted by humans. | 1. Road Conditions aspects: "*holes*", "*flood*", "*pavement*"<br>2. Delay Times aspects: "*slow*", "*stop*", "*jam*"<br>3. Tolls or Fares aspects: "*cheap*", "*costly*" |
| | | | | 2. Traffic but posted by robots. | |
| | | | | 3. Non-traffic. | |
| 2 | Italian | Twitter's API, Twitter4J | Yes | 1. Traffic. | "*traffico*" (traffic) or "*coda*" (queue) or "*incidente*" (crash) |
| | | | | 2. Non-traffic. | |
| | | | | 1. Traffic due to external event | 1. ("*traffico*" (traffic) or "*coda*" (queue)) and "*partita*"(match)<br>2. ("*traffico*" (traffic) or "*coda*" (queue)) and "*processione*" (procession)<br>3. ("*traffico*" (traffic) or "*coda*" (queue)) and "*concerto*" (concert)<br>4. ("*traffico*" (traffic) or "*coda*" (queue)) and "*manifestazione*" (demonstration) |
| | | | | 2. Traffic congestion or crash | 1. "*traffico*" (traffic) and "*incidente*" (crash)<br>2. "*traffico*" (traffic) and "*coda*" (queue)<br>3. "*incidente*" (crash) and "*coda*" (queue) |
| | | | | 3. Non-traffic. | |
| 3 | English | Twitter's API | Yes | 1. Traffic. | "vehicle", "accident", "road", "collision", "crash", "wreck", "injury", "fatal accident", "casualty". |
| | | | | 2. Non-traffic. | |
| 4 | English | Twitter's API | N/A | 1. Traffic. | e.g. "car accidents" |
| | | | | 2. Non-traffic. | |
| 5 | Thai | Twitter's API | N/A | 1. Traffic. | e.g. "รถติด" (traffic congestion) and "อุบัติเหตุ" (accident) |
| | | | | 2. Non-traffic. | |

Kokkinogenis et al. [1] extracted location using Named Entity Recognition (NER), and Entity Linking (EL) methods. NER is employed to extract references of locations from the tweets' messages. While EL is used to distinguish these references from each other as two streets may have the same name in different cities. However, NER and EL [15, 16] are more challenging in texts (e.g. tweets) which are informal and short. As a result, the authors got an unpleasant result (i.e. 24% F1-Score) as shown in TABLE II. Furthermore, they filtered the stop-words by removing words that did not provide useful information such as articles, conjunctions, and noisy words. This filter helps to minimize the impacts of none useful information in the analysis which may provide incorrect classification.

D'Andrea et al. [2] used five various efficient text mining techniques to have a good classification by formalizing the text in a good way. First, tokenization technique was employed in tweet texts to transform them into a stream (i.e. tokens) of syllables, words, and phrases. Second, the stop-word filtering method was used on the tokens to remove useless information. Third, the stemming technique was applied to transform each token to its root form by eliminating its suffix. Fourth, the stem filtering technique was employed to remove any stem which did not belong to the set of relevant traffic stems called "F relevant stems". The last method was called feature representation to assign a weight for every relevant stem or word in the feature vector. The F relevant stems and their weights were extracted and computed during the supervised learning stage according to their importance (i.e. frequent occurrences) in the dataset using the Inverse Document Frequency (IDF) index. These text mining techniques produced robust good results (e.g. 95.75% accuracy and 95.8% F1-score) for 2-class label classification and (e.g. 88.89% accuracy and 98.97% F1-score) for multi-class label classification. Whereas, multi-class label classification is more challengeable work since it requires to distinguish two kinds of information (i.e. traffic due to external event, and traffic congestion or crash event) in tweets related to traffic events.

Schulz et al. [3] applied text mining techniques in tweets to prepare the data and extract features. The preprocessing phase includes seven processes. First, the pre-filtering method was applied to remove duplicate tweet which did not contain any additional useful information, and removing "@-mentions" as were assumed to be not relevant to the traffic detection. Second, eliminating the stop-words technique was applied. Third, correcting spelling errors using Google Spellchecking API was employed. Fourth, the slang replacement process was performed to transform abbreviations to formal words by using www.noslang.com. Fifth, Stanford POS tagger filtering was applied using "POS" software to extract proper nouns from the tweets. Sixth, the temporal mention replacement process was employed to distinguish chronological expressions in order to eliminate the impact of overfitting the classification model. Lastly, the spatial mention replacement technique was employed to detect place and location.

Several feature extractions had been applied in the literature, where we will discuss only the features that gave the best results of this paper [3]. First, word-n-grams splitting was employed to divide a tweet text into a contiguous sequence of n words. For example, the word-3-grams of a tweet text like "To-day the traffic is indeed unbelievable" is "Today the queue" and "is indeed unbelievable". Second, the syntactic features were extracted, e.g., the number of capitalized characters, "!", and "?" in a tweet text. Finally, the accumulated TF-IDF score feature was extracted to measure the reflection of the importance of a tweet compared to all positive tweets in the training dataset. Therefore, using these proper text mining techniques for preprocessing and features' extractions helped give good results (i.e. 90.24% F1-Score).

Li et al. [4] extracted temporal and spatial information from the tweets and stored them in the database. They used several methods to extract features. First, a traffic URL feature was extracted such that if a tweet contained a URL to a news website then it might be related to traffic conditions. The second feature is the similarity of tweets to measure the frequency of a tweet with other similar tweets within time and geo-location ranges. The third feature is hashtag frequency which is the number of tweets that have the same hashtag used within a tweet. The last one is traffic-specific features to assign "Yes" value if the tweet contains time, number, and geolocation otherwise "No". These techniques gave a fair result (e.g. 80% accuracy) where this result was accumulative with all other kinds of detections including traffic events since this paper was a general-purpose detection.

Wanichayapong et al. [5] used different methodologies of preprocessing for the tweets. First, they created a special dictionary to categorize every traffic word into a place, verb, ban word, or prepositions. Then, every word in a tweet was tokenized and categorized using this dictionary. After that, a heuristic filtering system was employed to extract real traffic tweets. To be considered as traffic-related tweets, the tweets must contain a place and verb words according to this specific dictionary, not contain any ban word, and not have question words. Next, start and end attributes were extracted by using start and end preposition, with places from the tweet texts instead of geo-location inside the tweet's metadata. These techniques did not give robust good results (e.g. 75.7% F1-score). This was because extracting place from the tweet text might not be a proper way since the tweets were informal and might not refer accurately to the location of the events.

TABLE II.
CLASSIFIERS AND THEIR BEST RESULTS

| References | Class Label | Evaluation of Dataset | Classifiers | Best Results | | | |
|---|---|---|---|---|---|---|---|
| | | | | Accuracy | Precision | Recall | $F_1$-Score |
| 1 | 3 | Manual | SVM | N/A | 54.1% | 15.4% | 24.0% |
| 2 | 2 | 10-fold cross validation | SVM, C4.5, KNN, NB, PART | 95.75% (SVM) | 95.8% (SVM) | 96.25% (SVM) | 95.8% (SVM) |
| | 3 | | SVM, C4.5, KNN, NB, PART | 88.89% (SVM) | 89.6% (SVM) | 88.9% (SVM) | 88.97% (SVM) |
| 3 | 2 | 10-fold cross validation | SVM, NBB, JRip | 90.24% (SVM) | 90.4% (SVM) | 90.2% (SVM) | 90.2% (SVM) |
| 4 | 2 | N/A | Linear Regression Model | 80.00% (for all incidents) | N/A | N/A | N/A |
| 5 | 2 | N/A | Syntactic Analysis | 93.23% | 62.77% | 95.36% | 75.70% |

*C. Classification*

Kokkinogenis et al. [1], D'Andrea et al. [2], Schulz et al. [3] applied the SVM classifier which gave better results than other classifiers used in these three papers. SVM is a popular classifier which is defined by a separating hyper-plane. The output of this algorithm is the optimal hyperplane that categorizes new examples. This algorithm can handle high dimensional vector spaces, which makes feature selection less critical.

Kokkinogenis et al. [1] employed human users, and robot users to generate the training dataset with formal messages similar to tweet texts to easily classify it. Then, SVM was employed for classification. However, using a robot was not a good idea as real tweets were informal and short. This approach did not give good results (e.g. 24% F1-score).

SVM, C4.5, KNN, NB, and PART classification algorithms were employed by D'Andrea et al. [2]; whereas, SVM, NBB, JRip were employed by Schulz et al. [3]. Li et al. [4] applied Linear Regression Model. The syntactic analysis was used by

Wanichayapong et al. [5] to classify traffic tweets and specify geo-location according to the start and end attributes by using Google Geocoding API.

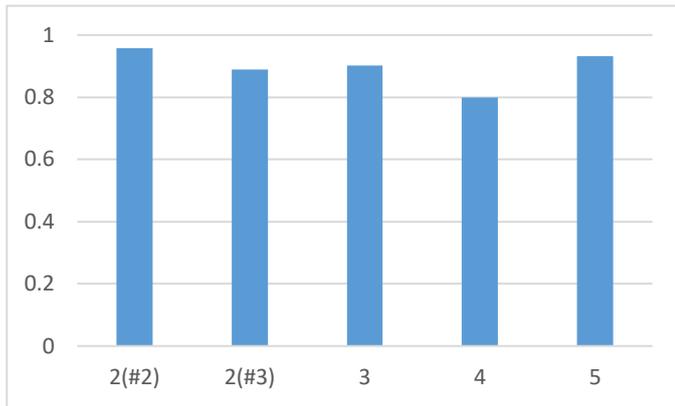

Fig. 1. The accuracy results.

### D. Evaluation

TABLE II shows the evaluation of the predictive models. Kokkinogenis et al. [1] used a manual process to evaluate its experiment. D'Andrea et al. [2] and Schulz et al. [3] employed 10-fold Cross-Validation to evaluate the classifiers. 10-fold cross-validation is commonly used to split the datasets. The datasets are randomly partitioned into 10 approximately equally sized subsamples. Subsequently, 10 iterations of training data and test data are performed such that within each iteration a fold of the data is used once for testing while the remaining 9 folds are used for training datasets. This approach is very important if the amount of data is small (e.g. small-scale events). This helps traffic events' experiments to reduce the sensitivity of the partitioning. Whereas, Li et al. [4] and Wanichayapong et al. [5] did not mention the way that they evaluated predictive models.

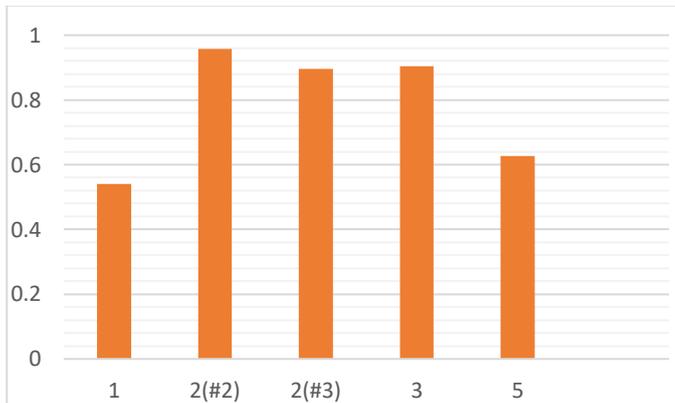

Fig. 2. The precision results.

Four different measurements (i.e. accuracy, precision, recall, and F1-score) were used to evaluate the correctness and performance of the classifiers [7]. These evaluations depend in four values, viz., True Positive "TP" (i.e. the number of real positive instances correctly classified as a positive class label), False Positive "FP" (i.e. the number of real positive instances incorrectly classified as a positive class label), True Negative "TN" (i.e. the number of negative instances correctly classified as a negative class label), and False negative "FN" (i.e. the number of negative instances incorrectly classified as a negative class label).

Accuracy was used to measure the accuracy of predicting the class label. The accuracy measure may not be well suited for evaluating models derived from an imbalanced dataset since the positive or negative class labels are high or low. The accuracy formula is as follows:

$$Accuracy = \frac{TP+TN}{TP+FP+FN+TN} \quad (1)$$

Precision (i.e. positive predictive value) was used to measure the ratio of relevant test instances, as the following formula:

$$Precision = \frac{TP}{TP+FP} \quad (2)$$

Recall (i.e. sensitivity) is the ability of the test to correctly detect positive instances, as the following formula:

$$Recall = \frac{TP}{TP+FN} \quad (3)$$

F1-score is a measurement that takes into consideration both precision and recall to make a good judgment about the results especially in an imbalanced dataset, as the following formula:

$$F_1 - score = \frac{2*Precision*Recall}{Precision+Recall} \quad (4)$$

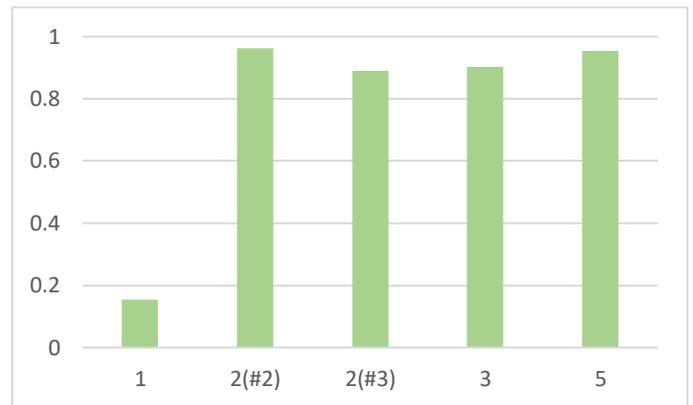

Fig. 3. The recall results.

### E. Analysis

In analyzing the performance of the best methods in all papers [1, 2, 3, 4, 5] using SVM classification method, linear regression model, and syntactic analysis over two and three class labels, we found that SVM method in paper [2] with 2-class label had delivered the best results of the four measurements as shown in Fig. 1, Fig. 2, Fig. 3, Fig. 4, and TABLE II. This might be related to that the authors followed the proper text mining techniques. This was by using useful information of the tweets to formulate a proper weight for every root word according to its occurrences in the training dataset. Also, they used a balanced training dataset in supervised learning to build the model. Furthermore, they manually checked the class labels of every tweet in the training dataset and found that 4% of tweets had the keywords of the positive class label but these tweets were actually negative instances. Therefore, they set the class label for these tweets as negative class label.

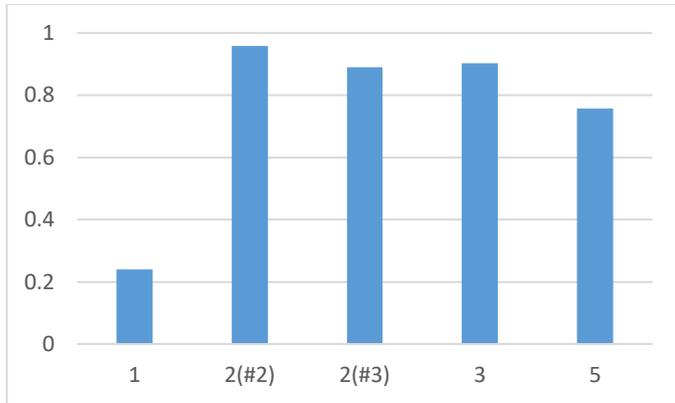

Fig. 4. The $F_1$-score results.

The most unpleasant results were in paper one [1] as shown in Fig. 2, Fig. 3, Fig. 4, and TABLE II. This was because the use of less efficient methods to extract features. For instances, NER and EL methods suffer from informal and short texts as in tweets [15, 16]. Also, the authors did not apply state-of-the-art techniques such as stem filtering and stem weighting.

However, none of the papers [1, 2, 3, 4, 5] can be reliable to detect traffic in real time for every event. We can see in paper [2] that sometimes the system delayed more than (50) minutes than the news channel. This delay was comparing to the news channel posting time rather than the real event time; that means, this late can be in fact worse. Thus, we believe that this field of research requires more effort to get accurate traffic detection system in real time. This may be achieved by using other micro-blogging services like Facebook; this is because Facebook is more popular than Twitter.

### III. CONCLUSION

This survey paper focused in deliberate a summary view of current state-of-art in traffic detection events. we used the latest related papers [1, 2, 3, 4, 5] in the traffic detection field. The goal of these papers was to build ITSs which were a real-time detection of traffic-related events from Twitter stream. They might notify the users about the presence of traffic events. These papers used many text mining and machine learning techniques to classify tweets. Some authors employed the state-of-the-art techniques like D'Andrea et al [2], in order to detect traffic events *often* before online news web sites and local newspapers. Moreover, the results in the second paper [2] was the best among all other papers (i.e. 95.75% accuracy and 95.8% $F_1$-score) using SVM classification algorithm. However, the big issue of the current research was developing a detection system using social network in real-time.

### IV. FUTURE WORK

We may see a better detection system in real time in micro-blogging services in future. This may be achieved by using other micro-blogging services like Facebook. Facebook is more popular than Twitter. The second paper used Italian language. Applying the methodologies used in this paper in USA states for SUMs (e.g. Facebook wall posts) in English language may provide a good traffic detection system in real-time.

### ACKNOWLEDGMENT

I thank Professor Wen Xu of New Mexico State University for helping and supporting me in social computing techniques and methodologies. I also thank all my classmates in C S 579 Special Topics (Social Computing) in New Mexico State University that I have learned more interesting topics and expanded my knowledge from their presentations and projects during the Fall 2016 semester.